\documentclass[prb,aps,twocolumn,showpacs,amsmath,amssymb,floatfix,eqsecnum]{revtex4}
\usepackage{graphicx,amssymb,natbib,epsfig}
\usepackage{verbatim}

\begin{document}

\def\OMIT#1 {}
\def\LATER#1 {{\bf LATER  #1 }}
\def\MEMO#1 {{\bf #1 }}
\def\NOTE#1 {{\bf #1 }}
\def\SAVE#1 {}

\newcommand{\eqr}[1]{(\ref{#1})}
\newcommand{\half}{{\textstyle{\frac{1}{2}}}}
\newcommand{\HH}{{\cal H}}
\newcommand{\Sop}{{\cal S}}
\newcommand{\Oop}{{\cal O}}
\newcommand{\HHO}{{\HH^{0}}}
\newcommand{\ti}{\tilde}
\newcommand{\pa}{\parallel}
\newcommand{\HHop}{{\cal H}^{\rm hop}}
\newcommand{\VU}{V}
\newcommand{\HVU}{{{\cal H}^V}}
\newcommand{\HVUbox}{{\HVU_\Box}}
\newcommand{\Hinter}{{\cal H'}}
\newcommand{\Hsigma}{{{\cal H}_\sigma}}
\newcommand{\la}{\langle}
\newcommand{\ra}{\rangle}
\newcommand{\Proj}{{\hat{\mathcal{P}}}}
\renewcommand{\SS}{{\bf S}}
\newcommand{\Sz}{{S^z}}
\newcommand{\Sx}{{S^x}}
\newcommand{\Sy}{{S^y}}
\newcommand{\TT}{{\bf T}}
\newcommand{\Tz}{{T^z}}
\newcommand{\Pplus}{{P^+}}
\newcommand{\Pminus}{{P^-}}
\newcommand{\dagg}{^\dagger}
\newcommand{\rr}{{\bf r}}
\newcommand{\Nbox}{{n^{\Box}}}
\newcommand{\nbox}{{n^{\Box}}}
\newcommand{\tc}{\tilde{c}}
\newcommand{\be}{\begin{equation}}
\newcommand{\eqend}{\end{equation}}

\title{SPONTANEOUS CURRENTS IN SPINLESS FERMION LATTICE MODELS AT THE  STRONG-COUPLING LIMIT}

\author{Sumiran Pujari and C. L. Henley}

\affiliation{Department of Physics,
Cornell University, Ithaca, New York 14853-2501}

\begin{abstract}
What kind of lattice Hamiltonian manifestly has an ordered
state with spontaneous orbital currents?
We consider interacting spinless fermions on an array of
square plaquettes, connected by weak hopping; the array
geometry may be a $2\times 2L$ ladder, a $2\times 2 \times 2L$
``tube'', or a $2L \times 2L$ square grid.  At half filling,
we derive an effective Hamiltonian in terms of pseudospins,
of which one component represents orbital currents, and
find the conditions sufficient for orbital current long-range order. 
We consider spinfull
variants of the aforesaid spinless models and make contact with
other spinfull models in the literature purported to possess
spontaneous currents.
\end{abstract}



\maketitle


\section{Introduction}

In condensed matter physics, strongly correlated electrons underly a 
great variety of ordered states, both common and exotic 
(e.g. ferromagnets, superconductors). One of the lesser-studied orders
is spontaneous currents (known sometimes as ``orbital antiferromagnetism'').
In this paper, we seek a {\it minimal} (spinless) 
toy model that manifestly exhibits such currents,
precisely because any systematic study of fermion orderings with a 
quadratic order parameter reveals that the possible ordered states
include not only the familiar cases 
of charge or spin density waves or superconductivity,  but also 
spontaneous orbital currents~\cite{halperin-rice,Schulz,varma,nayak,ddw}. 
Yet such states have not been definitively observed in any material,
nor numerically in the Hubbard model~\cite{jarrell}, and only very
recently for any realistic microscopic Hamiltonian~\cite{weber-giamarchi,new-giamarchi}.
Thus, we ask : Which aspects of the interactions 
and/or degrees of freedom dispose a system generically 
towards ordered states with spontaneous currents ?

Such states were considered especially in the context
of high-$T_c$ cuprates.  Early in their history,
``flux phases'' with current order were 
invented~\cite{Aff-Mars,Dombre-Kotliar,ivanov}
however the actual phase was expected to be disordered.
More recently,
two different kinds of spontaneous-current order were 
advanced to explain the mysterious pseudogap state of high-$T_c$ 
cuprates\cite{varma,ddw}. Ref.~\onlinecite{ddw} proposed the 
``d-density wave'', which  breaks translational symmetry 
(currents circulate in  opposite senses around even and odd plaquettes);
variants were considered more recently~\cite{ivanov-new},
e.g. modulated versions~\cite{weber,raczkowski}.

In contrast, Varma's phases~\cite{varma,simon-varma} require 
the so-called ``three band'' model 
in which oxygen orbitals of the CuO$_2$ layer 
are explicit independent degrees of freedom; 
the latter state breaks 4-fold rotational and time-reversal symmetry, 
but not translational symmetry. 
Experiments on photoemission~\cite{compuzano} (in BSCCO) 
and neutron diffraction~\cite{bourges} (in YBCO) 
indicated time-reversal symmetry breaking, 
in the pattern of Ref.~\onlinecite{simon-varma}.
Finally, Khomski and collaborators showed
currents are implied by non-coplanar spin order in 
(spinfull!) Mott insulators~\cite{khomskii-currents}.

\SAVE{Early papers~\cite{Dombre-Kotliar,Aff-Mars}
constructed spontaneous-current states as Gutzwiller projections 
of the free-electron ground state with an alternating pattern of
fluxes threading the plaquettes; this was meant as a variational state 
for a (single-band) Hubbard model, if the doped antiferromagnet
behaves as a doped spin liquid.    But the long-range current order
was assumed to be a spurious feature (compare Ref.~\onlinecite{ivanov})}

These proposals motivate a basic question:
under what circumstances, in principle, can a quantum state be realized 
with spontaneous currents? Where, in a model's parameter space, is such 
a state favored? 
Ever since the Hubbard model, toy lattice models having 
a minimal parameter space (and possibly amenable to solution)  
have been key tools to sort out basic questions such as these.
For the more familiar orders, ``strong-coupling'' models are well-known 
in which some ``zero-order'' state trivially has the order in question, 
and the order is stable against small perturbations. Thus, in the phase 
diagram, one is assured of a corner where the ordered  phase occurs and 
extends an indetermined distance towards the regime where perturbations 
are large (which is usually the physical regime).  But in the case of 
currents order, no general intuitive picture has emerged.

This paper addresses this question using a toy-model 
built from square plaquettes; focusing mainly
on the simplest case of {\it spinless} 
fermions, we explore the possibilities for realizing spontaneous currents. 
The main prior study of orbital currents in spinless models is
Nersesyan's ladder model~\cite{nersesyan,narozhny}, in which a map to 
spinfull chains was introduced that we adopt in Sec.~\ref{sec:nersesyan}.
Quite recently, spinless models were motivated by the possible
realization in cold dilute atoms~\cite{kolezhuk}.


The choice of square plaquettes is a choice motivated both by convenience 
of calculation and real material geometries. 
As we will see, a square plaquette has spontaneous currents as one of 
its natural degrees of freedom which is what we desire to investigate: 
possibility of spontaneous currents in the zero-order ground state.

This paper is organized as follows: 
In Section \ref{sec:setup}, we define our toy-model Hamiltonian and 
set up the various lattice geometries -- tube, ladder, and square lattice
--  we shall deal with;  we go on to describe the properties of 
one square plaquette as it forms the basic unit 
of all the lattice geometries considered, in particular
reducing its degrees of freedom to a pseudospin 
via the method of canonical transformations (which is briefly 
summarised in the appendix, as it is the basis of all our subsequent calculations.) 
The core section is Sec.~\ref{sec:effham}, where we implement
the pseudospin projection
(illustrating it in detail for the case 
of a ``tube" lattice) and obtaining a pseudo-spin effective Hamiltonian,
showing its final form for the respective lattices; we
also explore the relation between
the fermion Hamiltonian and the pseudospin Hamiltonian,
focusing on possibility of spontaneous currents in the ground state.
In Sec.~\ref{sec:spinfull}, we connect out work to spinfull models
in two ways: simply incorporating spin (Sec.~\ref{sec:addspin}) or
mapping a pair of site indices to spin labels (Sec.~\ref{sec:nersesyan}).
At last in Sec.~\ref{sec:uniform} we ask if we have learned how to
construct a uniform lattice model with currents.
We conclude (Sec.~\ref{sec:discussion}) by discussing why it is hard to
obtain spontaneous current order, and what light this may shed on 
realistic motivated models of such order.

\section{Microscopic Model and pseudospin mapping}
\label{sec:setup}

Our basic model Hamiltonian is $\HH =  \HHop + \HVU + \Hinter$ with
  \begin{subequations}
  \label{eq:Ham}
  \begin{eqnarray}
       \HHop   &\equiv& -t \sum _{{\rm n.n.}\Box} [c\dagg(\rr)c(\rr') + h.c.] 
       \label{eq:Ham-t}\\
       \HVU     &\equiv& + \VU \sum _\Box\hat n(\rr) n(\rr')    
       \label{eq:Ham-V}\\
       \Hinter &\equiv& -t' \sum _{{\rm n.n.}\Box-\Box} 
                    [c\dagg(\rr)c(\rr') + h.c.].  
        \label{eq:Ham-tprime}
  \end{eqnarray}
  \end{subequations}
Each site $\rr$ has an orbital with room for one spinless fermion.
A disjoint set of plaquettes (``strong plaquettes'') are singled out. 
Within each strong plaquette (tagged by ``$\Box$'' in notations)  
there is a hopping $-t$ on every bond; there is also a repulsion $\VU$ between 
{\it any} two fermions (whether first or second neighbors; in a spinless
model, of course, there can be no onsite term).

\begin{figure}[h]
\includegraphics[width=1\linewidth]{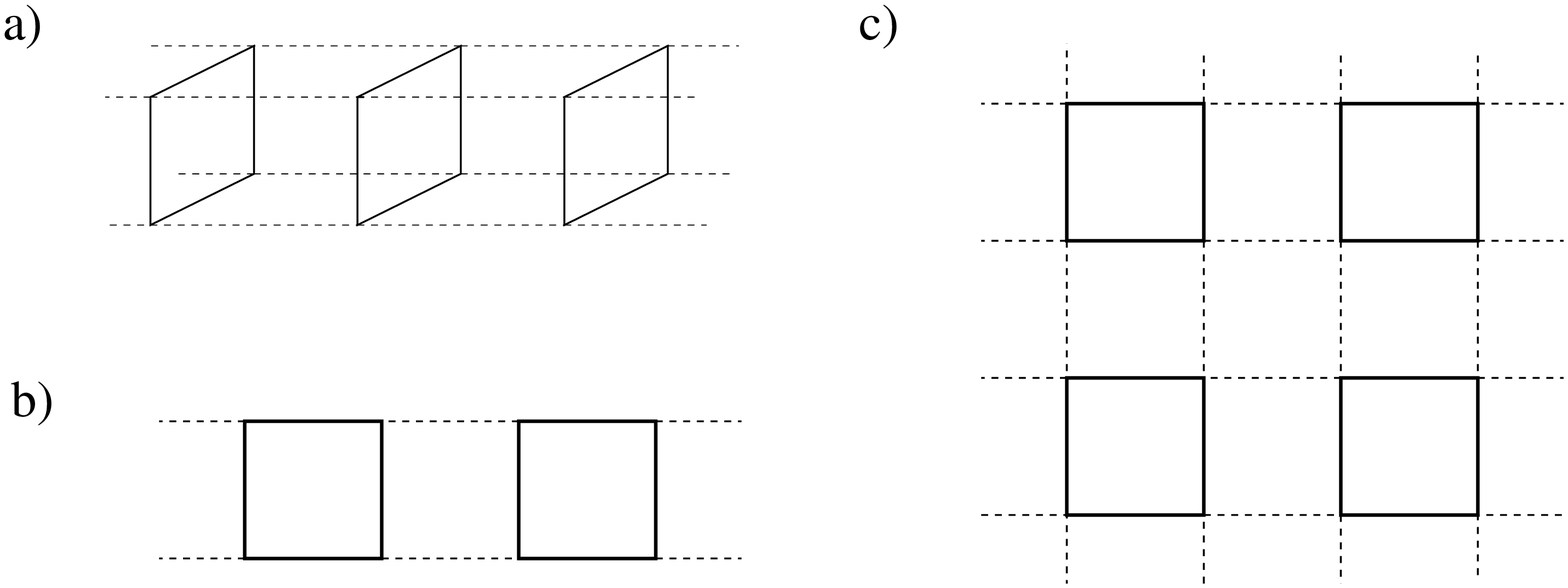}
\caption{Geometry of a (a) Ladder, (b) Square Lattice and (c) Tube. 
The fermions reside on the vertices of the lattices shown. 
}
\label{fig:geometries}
\end{figure}

Finally, every bond between bold plaquettes has a hopping $-t'$,
which is assumed to be a small perturbation. 
We will consider three kinds of lattice geometries,
as shown in Fig. \ref{fig:geometries}.
(a) a ladder, in which every other plaquette is strong;
(b) a ``tube'', which is one-dimensional like the  ladder,
but the strong plaquettes are oriented transverse; and 
(c) a square lattice, in which one of four plaquettes is strong.  
The ladder is simplest, but also has the least symmetry.

We were guided by three considerations while concocting 
the model of \eqr{eq:Ham}.
Firstly, we desired the spontaneous currents to be 
\emph{explicitly} related to the degrees of freedom 
describing our (degenerate) ground state subspace.
(See subsection~\ref{sec:pseudospin}, below.)
Secondly, the (zero temperature) behavior should be obvious in a strong coupling limit.
A standard trick~\cite{Mila,Tsai_Kivelson}.
to achieve both ends is to artificially weaken some bonds
thereby introducing a small parameter ($t'$ in our case).
In the $t'=0$ limit, the system decomposes into
small disjoint clusters, each with a degenerate ground state whose 
operators are represented by pseudospins;.
As the small parameter is perturbatively turned on, it generates
an effective Hamiltonian between the pseudospins; from the symmetry 
of the effective Hamiltonian, one can often read off the
symmetry of its ground state.

Finally, to make our model more physical, 
we limit the terms to fermion hoppings and interactions
 and no other four-fermion terms.  Also, as we hope that
our model(s) might later be adiabatically connected to a
uniform one (see Sec~\ref{sec:uniform}),
if a certain term is included (say) within strong plaquettes,
we will be open to including an inter-plaquette term of the
same form (with arbitrary -- small -- coefficients). But we never assume 
any particular condition on the ratios between the intra- and inter-plaquette
terms, except that all of the latter are small for perturbation purposes.


\subsection{Eigenstates of Disconnected Plaquettes}
\label{sec:oneplaq}

Let $\HH_0$ include the $\HHop$ and $\HVU$ terms, representing
a set of disconnected squares.  
We will work at half filling, i.e. two fermions per square on average,
but our Hilbert space includes all ways of distributing these
over the plaquettes.

Consider an isolated strong plaquette, with sites $x=0,1,2,3$
forming a ring.
Note $\HVUbox$ is the same for all states
accessible by hopping, so if there are $\Nbox$ fermions on the plaquette,
$ \HVUbox = \half \Nbox(\Nbox-1) \VU$ drops out like a $c$-number:
as in a noninteracting model~\cite{FN-Vterm},
multi-fermion states are built from the one-particle
eigenstates on the ring, defined by creation operators
   \be
  \tc\dagg_m \equiv \half \sum _x e^{\half i \pi m x} c\dagg(x)
   \label{eq:mstates}
   \eqend
where $m=0,\pm 1, 2$ is the angular momentum around the ring.
The single-fermion eigenenergies are 
    \be
    E_m=-2t \cos(\half \pi m),
   \eqend
i.e., $E_0=-2t$, $E_{\pm 1}=0, E_{2}=+2t$.
Table \ref{tab:E-plaquette} lists the multi-fermion
ground states for each occupation sector of a single plaquette.
Our interest will be the 2-fermion sector since it
has degenerate ground states $|2 +\ra$ and
$|2 -\ra$ with spontaneous current in the  
$+$ and $-$ senses, respectively.

\SAVE{I moved this para up before the subsection heading -- YES}
To have any possibility of a symmetry broken state, (at least some
of) the plaquettes must be in the degenerate half-filled ground states.
What is the ground state of an extended system of $N$ sites forming
$N/4$ disconnected plaquettes with $N/2$ fermions (i.e. half filling)?
The case $\VU=0$ is more degenerate than we wished, since any  combination of 
states with $\Nbox=1,2,3$ has total energy $-2t(N/4)$.  However, taking $\VU>0$
favors the subspace in which $\Nbox=2$ on every plaquette.
In that case, the only freedom is the senses of the currents in each
of the $N/4$ plaquettes, giving a degeneracy $2^{N/4}$.

\begin{table}
\caption{States with $\Nbox$ fermions on a plaquette.}
\begin{tabular}{|llll|}
\hline
$\Nbox$ & label & occupation~~~  & energy \\
\hline
 0   &   --       &  --                  &  0  \\
 1   &  $|1\ra$   & $|\ti{0}\ra$           &   $-2t$ \\
 2   &  $|2+\ra$  & $|\ti{0},+\ti{1}\ra$ & $-2t+\VU$ \\
     &  $|2-\ra$  & $|\ti{0},-\ti{1}\ra$ & $-2t+\VU$ \\
 3   &  $|3\ra$   & $|\ti{0},+\ti{1},-\ti{1}\ra$ & $-2t+3\VU$ \\
 4   &  $|4\ra$   & $|\ti{0},+\ti{1},-\ti{1}, \ti{2}\ra$ & $6\VU$ \\
\hline
\end{tabular}
\label{tab:E-plaquette}
\end{table}

\subsection{Pseudospin Mapping.}
\label{sec:pseudospin}

These states can be labeled as an array of spin-$\frac{1}{2}$ pseudospins
$\vec{P}_\alpha$ with $P^z_\alpha =\pm 1/2$ when plaquette $\alpha$ is in
state $|2\pm\ra$.  We aim, via second-order perturbation in
$t'$, to compute the effective Hamiltonian $\Hsigma$ defined within
the ground state manifold (and thus taking the form of a spin
Hamiltonian in $\{ \vec{P}_\alpha \}$.)

The spin-$\frac{1}{2}$ pseudospin Hilbert space can be defined as follows :
\begin{eqnarray}
| \pm \ra_z & \equiv & |\pm\ra \equiv |2 \pm \ra \nonumber \\
| \pm \ra_x & \equiv & \frac{1}{\sqrt{2}}
   \left(|2 \pm \ra \pm |2 \mp \ra\right) \nonumber \\
| \pm \ra_y & \equiv & \frac{1}{\sqrt{2}}
   \left(|2 \pm \ra \pm i|2 \mp \ra\right) \nonumber \\
\label{eq:pseudospin-basis}
\end{eqnarray}

\begin{figure}[ht]
\centering
\includegraphics[width=0.9\linewidth]{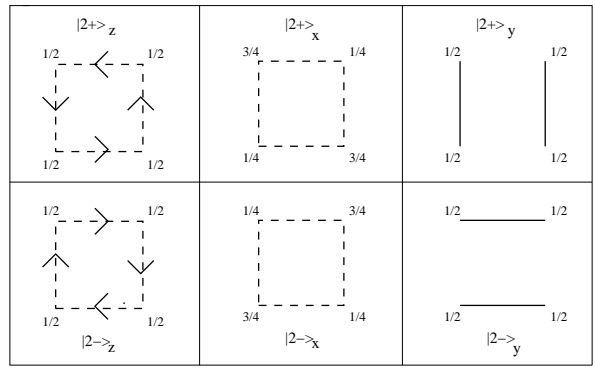} 
\caption{Properties of single plaquette eigenstates. Arrows represent orbital currents. Lines represent the bond density;
non-dashed line represents unit bond density, dashed line represents half and absence of line represents zero bond density.
The numbers at the corners of the plaquette represent the charge density at the respective sites. Note that they add up to two
corresponding to half-filling.}
\label{fig:eigen}
\end{figure}

Different orders of the fermions -- spontaneous currents, and (site- or bond-centered) charge density waves -- correspond to expectations of
three characteristic operators; when projected to the pseudospin subspace \eqr{eq:pseudospin-basis}, these reduce to the
three pseudospin operators (here $i,j=0,...,3$ label sites
counterclockwise around a plaquette,  as in Figs.~\ref{fig:calctube} and
\ref{fig:calcladder}):

{1. \em{Pseudocurrent operator:}}
\begin{subequations}
\label{eq:oper_map}
\begin{gather}
      \hat{I}_{ij}  
      = - \hat{I}_{ji}  
\equiv  { i(c\dagg_i c_j - c\dagg_j c_i) }  \to  { \frac{\hat{P}^z}{2} } 
       \label{eq:Sz} \\
\intertext{2. \em{Charge Density operator:}}
      \hat{n_i}  \equiv  { c\dagg_i c_i }  
             \to { (-1)^i\frac{\hat{P}^x}{2}  + \frac{1}{2} } 
   \label{eq:Sx}  \\
\intertext{3. \em{Bond Density operator:}}
      \hat{B}_{ij}  \equiv   { (c\dagg_i c_j + c\dagg_j c_i) }  \to  {-(-1)^{(i+j)} \frac{\hat{P}^y}{2} + \frac{1}{2} } 
   \label{eq:Sy}
\end{gather}
\end{subequations}
Here ``$\to$'' means the operators have the same matrix elements
when acting in the pseudospin Hilbert space.
Any operators in the pseudospin subspace of a plaquette can be expressed in terms of 
$\vec{P} \equiv (\hat{P}^x, \hat{P}^y, \hat{P}^z)$.
Fig. \ref{fig:eigen} depicts states in which the respective
operators have expectations.

So, for purposes of nomenclature, we call the pseudospin states in the $z$-direction 
as current carrying states (which is expected since $+\ti1$ and $-\ti1$ carry momentum) 
or CCS, the pseudospin states in the $x$-direction as charge density waves or CDW, 
and the pseudospin states in the $y$-direction as bond density waves or BDW. Spontaneous
currents, Orbital currents or just currents will be used interchangeably to refer to
CCS as they have been used in the literature before.
(The ``bond order'', making different directions inequivalent
without a translational modulation of the charge density,
would be an  example of ``electron nematic''~\cite{electron-nematic,kim}
if all similarly oriented bonds had the same order parameter.)

Incidentally, we use the term {\it pseudo}current operator because this is
not the true current operator.  The latter would be the time derivative of the 
charge density operator, and is evaluated as a commutator of the charge density operator 
with the full Hamiltonian. Most often, the true current is proportional to 
(or at least has overlap with) the pseudocurrent operator; then, any state 
with pseudocurrent order will also have true current order.
(The pseudocurrent would be the real current if the
Hamiltonian contained only nearest-neighbor hopping.)

\section{Effective Pseudospin Hamiltonian}
\label{sec:effham}

In this section, we go on
to calculate an effective Hamiltonian by 
second-order perturbation theory, formulated via canonical
transformations (reviewed briefly in Appendix~\ref{app:canonical}.).
We shall consider several variations on the model Hamiltonian
\eqr{eq:Ham}; the result is always a special case of the
general form
   \begin{multline}
      H_\sigma = \sum _{\la \alpha\beta\ra}
\left[ J_x P^x_\alpha P^x_\beta + 
 J_y  P^y_\alpha P^y_\beta  +
 J_z  P^z_\alpha P^z_\beta \right] \\
  - \sum _\alpha 
\left[ h_x P^x_\alpha  + 
       h_y P^y_\alpha  + 
       h_z P^z_\alpha \right] .
   \label{eq:Ham_spin}
   \end{multline}
Here $\vec{P} _\alpha$ is the pseudospin; 
($\alpha$ runs over all strong plaquettes and $\la \alpha \beta \ra$
are nearest neighbors (in the ladder, tube, or square lattice arrangements).
In view of the mapping of operators \eqr{eq:oper_map}, the original system has
spontaneous-current order if and only if \eqr{eq:Ham_spin}
has pseudospin order in the $z$ direction.  Consequently, our
central concern is whether and how the terms in \eqr{eq:Ham_spin}
break pseudospin rotation symmetry.

The above form of the effective Hamiltonian is governed by the two different
kinds of inter-plaquette hopping processes that can occur at second-order perturbation theory.
As we will see in the next subsection, at second order, only two adjoining
plaquettes can take part in the hopping processes.
On the one hand, a ``degenerate'' hop takes a fermion from a plaquette
to a single-fermion orbital on the
adjoining plaquette, degenerate with the orbital it hopped 
out of; such hops are responsible
for all the pseudospin exchange interactions. 

On the other hand, an ``excited''  hop takes the fermion from
a partly-occupied degenerate orbital on one plaquette to a higher orbital
on the adjoining plaquette, which would not be occupied in any single plaquette
state; this kind also 
includes the case (related by particle-hole symmetry) of an electron
hopping {\it out} of a deeper orbital into a degenerate orbital on the
new plaquette, or even into a higher one  (and of course the reverse hops).
An excited-state hop is not conditioned on which state the plaquette
(with the excited state) was in: within the pseudospin manifold, that
only differs in the ``degenerate'' partly-occupied levels. 
Morever, the only way to return to the pseudospin manifold is to undo 
the same hop thus making exchange of pseudospins  impossible.
Consequently, excited-state hops can (at most) generate only single-pseudospin
terms in the effective Hamiltonian.

We shall first consider the tube model case (Sec.~\ref{sec:Heff_tube}),
since it has the greatest symmetry (the combination of two adjoining 
plaquettes has a 4-fold rotation). The other two cases 
(Sec.~\ref{sec:Heff_laddersq}) are variations on the tube case,
in that either additional terms appear (due to reduced symmetry)
or are accidentally canceled.

\subsection{Effective Hamiltonian for the Tube}
\label{sec:Heff_tube}

The perturbation ($t'$ hopping) changes the filling on
two plaquettes, hence no first order process stays in
the reduced Hilbert space (of $\nbox_\alpha=2$ on all plaquettes).
To do that in a second order process,
a fermion hops from one plaquette (``A'') 
to a neighboring one (``B''), and then a fermion
hops back from the second to the first plaquette.

\begin{figure}[ht]
\includegraphics[width=0.5\linewidth]{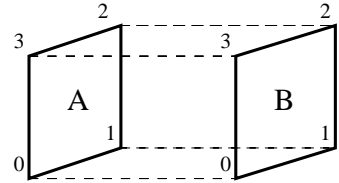}
\caption{The two-plaquette unit involved in a 
second order hopping process.  Site labels show our
convention for the (bipartite) ladder or square lattices.
}
\label{fig:calctube}
\end{figure}

For a pair of plaquettes on a tube (Fig. \ref{fig:calctube}), 
the perturbation takes the form
  \begin{eqnarray}
  \label{eq:tube-hop-pert}
      \Hinter & = & -t'(\sum_i c\dagg_{A,i} c_{B,i} + h.c.) \nonumber \\
		& = & -t'(\sum_m \tilde{c}\dagg_{A,m} \tilde{c}_{B,m} + h.c.)
  \end{eqnarray}
Notice that the hopping conserves the angular momentum around the plaquette. For this reason, the only excited states that can 
participate in the second order hopping processes are $|1;3\ra \equiv |\tilde{0};\tilde{0},+\tilde{1},-\tilde{1}\ra$ and 
$|3;1\ra \equiv |\tilde{0},+\tilde{1},-\tilde{1};\tilde{0}\ra$ and the corresponding non-zero matrix elements are 
\begin{eqnarray*}
\la1;3|\Hinter|2+;2-\ra & = & t' \nonumber \\
\la3;1|\Hinter|2+;2-\ra & = & t' \nonumber \\
\la1;3|\Hinter|2-;2+\ra & = & -t' \nonumber \\
\la3;1|\Hinter|2-;2+\ra & = & -t' \nonumber 
\end{eqnarray*} 
The rest of the matrix elements are zero. Thus, using Eq. \ref{eq:Cantran}, we get the following second order effective two-plaquette Hamiltonian
\begin{align}
   H_{tube} = & -\frac{2t'^2}{V} \Big( |2+;2-\ra \la 2+;2-| + |2-;2+\ra \la 2-;2+| \Big) \nonumber \\
          &  +\frac{2t'^2}{V} \Big( |2+;2-\ra \la 2-;2+| + |2-;2+\ra \la 2+;2-| \Big) \nonumber \\
   \label{eq:Htube}
    \end{align}
\textbf{Conversion to spin Hamiltonian}: 
In accord with our pseudospin mapping, we abbreviate $|2\pm\ra$  by $|\pm\ra$ to label the pseudospin states. Now, the transcription 
to spin notation (for pseudospin($P$)) is:
  \begin{eqnarray}
|2+\ra\la 2+| &\to &  \left(\half + P^z\right); \nonumber \\
|2-\ra\la 2-| &\to & \left(\half - P^z\right);\nonumber \\
|2+\ra\la 2-| &\to & \Pplus;\nonumber \\
|2-\ra\la 2+| &\to & \Pminus.\nonumber \\
\label{eq:spintran}
   \end{eqnarray}
Inserting Eq. \ref{eq:spintran} into Eq. \ref{eq:Htube}, we get for the infinite tube 
  \be
     H_{tube} = \sum_\alpha \frac{4t'^2}{V} 
\left[\frac{1}{4} + \vec{P}_\alpha \cdot \vec{P}_{\alpha+1} \right].
  \eqend
Thus, the effective pseudospin Hamiltonian for the tube is a one dimensional spin-$1/2$ Heisenberg Antiferromagnet 
which, as is well known, does not exhibit long range order 
but only power-law correlations.  

This calculation is not only reminiscent of, but
completely analogous to, the derivation of the effective Heisenberg antiferromagnetic
exchange interaction in a half-filled Hubbard model; the role of spin is taken
by our angular momentum, since it is conserved by the hopping along the tube.
Hence only our $|\pm \tilde{1}\ra$ single-particle states (analogous to 
spin up and spin down electrons)  take part in the second-order process, 
thus giving rise to effective pseudospin exchange of exactly the
same (rotationally symmetric) form as spin exchange in the Hubbard model. This
is exactly the content of the discussion on ``degenerate" hops in Sec. \ref{sec:effham}.

In other cases of our model (ladder or square lattice), 
the perturbation need not conserve angular momentum,
so excited states like
$|\tilde{0};\tilde{0},+ \tilde{1},\tilde{2}\ra$ 
or $|\pm \tilde{1};\tilde{0},+ \tilde{1},-\tilde{1}\ra$ 
may then mediate second-order processes via the ``excited-state'' hops 
defined at the beginning of this section
Thus, they will give rise to single-site pseudospin terms only.

\subsection{Ladder and Square Lattice}
\label{sec:Heff_laddersq}

\begin{figure}[h]
\includegraphics[width=0.5\linewidth]{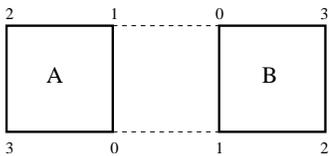}
\caption{The two-plaquette unit involved in a second order hopping process for the ladder. Site labels are shown, as used in e.g.
\eqr{eq:tube-hop-pert}.}
\label{fig:calcladder}
\end{figure}

In this sub-section, we list down the results of similar calculations for Ladder and Square Lattice cases. For the 
Ladder, the effective pseudospin Hamiltonian is 
\begin{align}
H_{ladder}  = & \sum_{\alpha} -\frac{t'^2}{V} \left[{P^z_\alpha P^z_{\alpha+1} - \frac{1}{2} (P^+_\alpha P^+_{\alpha+1} 
		+ P^-_\alpha P^-_{\alpha+1})}\right]  \nonumber \\
	      & + \frac{t'^2}{2V}\Big(\frac{1}{1+x}\Big) (P^y_\alpha) + const
\label{eq:PLadder}
\end{align}
where $x \equiv 2t/V$. To make the symmetry of the above expression clear, we make a simple transformation as follows.

\textbf{Staggered Pseudospins}: 
Let us define a new set of staggered spin operators $\vec{T}_\alpha$, by switching the definitions of ``up'' and ``down'' 
pseudospin on every other site ``B" by a $180^\circ$ rotation around $y$-axis. Then,
\begin{eqnarray}
\label{eq:stagger}
 P^z_B & \to & -T^z_B; \nonumber \\
 P^x_B & \to & -T^x_B; \nonumber \\ 
 P^y_B & \to & T^y_B; \nonumber \\
 P^{\pm}_A & \to & T^{\mp}_B 
\end{eqnarray}
while pseudospin operators on sites ``A" stay unchanged. 
This transformation converts Eq.~\eqr{eq:PLadder} to 
\begin{equation}
 H_\sigma = \sum_{\alpha} \left[ \frac{t'^2}{V} \left( const(x) + \vec{T}_{\alpha} \cdot \vec{T}_{\alpha+1}\right)+
    \frac{t'^2}{2V(1+x)} T^y_{\alpha}\right]
 \label{eq:TLadder}
 \end{equation}

The need for staggering the pseudospin arose out of a technicality. 
Let's focus on the effect of one pair of hops connecting
two plaquettes: e.g. 1--1 and 2--2 in Fig. \ref{fig:calcladder},
and compare it to the pair of hops 1--0 and 0--1 in 
Fig. \ref{fig:calctube}. 
(Indeed, to make the tube into a ladder,
we could cut the inter-plaquette tube bonds 0--0 and 3--3, 
alternating with cutting 1--1 and 2--2, down the line, then
flattening it out like an accordion-fold.)
Most importantly, looking in parallel directions along the
adjoining edges, the numbering around the B plaquette 
has the {\it opposite} sense 
from that of Fig. \ref{fig:calcladder}. 
We chose to use the numbering scheme shown so as to keep the 
clock sense the same on both plaquettes.

\SAVE{Sumi notes: the explanation of staggering, above, might be 
written better...after 2-3 readings it makes sense though.}

\SAVE{
We could have alternatively chosen the 
latter numbering scheme which would have eliminated the need for staggering but the clock sense would be reversed on alternate plaquettes.
The staggered pseudospins defined above correspond to the ``non-staggered" pseudospin definition for plaquettes with alternating
clock senses.}

Thus common sense combined with the pseudospin dictionary
\eqr{eq:oper_map} explains why the $x$ and $z$ components
flip sign, but the $y$ component does not, in \eqr{eq:stagger}.
For example, imagine an interplaquette interaction that favors 
having fermions at both ends of a weak bond
(see Sec.~\ref{sec:anisos-int}, below). We see in that 
for the tube (Fig.~\ref{fig:calctube})
that means (say) fermions take even sites on both
plaquettes,  but the same thing on the ladder 
(Fig.~\ref{fig:calcladder}) means they are even on one
plaquette and odd on the other; the difference between 
even and odd charge-order is a sign flip of the $P_x$ or $T_x$
component.  Also, the sign of pseudocurrents  (and hence
of $P_z$ or $T_z$) is manifestly flipped when the sense
around the plaquette is reversed.  On the other hand, 
if an interaction means that (say) bond-order on the 
1--2 bond of one plaquette of the tube (in Fig.~\ref{fig:calctube})
repels bond order on the 1--2 bond facing it, i.e. favored
antiferromagnetic alignment of the $P_y$ or $T_y$ components,
the same thing is true for the 0--1 bonds on the
ladder plaquettes (in Fig.~\ref{fig:calcladder}).


In Eq.~\eqr{eq:TLadder}, there is a uniform magnetic field in the pseudospin $y$
direction. We get the single-site terms from the ``excited" hops 
(defined at the start of this section) which are not disallowed for the ladder.
This competes with the antiferromagnetic exchange term, 
having the effect (as usual in antiferromagnets) of a uniaxial anisotropy 
favoring the $xz$ plane.  Hence, the system has the symmetry of an XY  model
ordering in that plane, which corresponds [by Eq.~\eqr{eq:oper_map}]
to CDW and spontaneous currents.  Similar to the tube, having a continuous symmetry
in one dimension, it would only have power-law correlations.

Doing the same for the square lattice amounts to extending the result of the ladder calculation to a square lattice. 
Recalling for the ladder [Eq. \eqr{eq:TLadder}], 
the pseudospin Hamiltonian for the two-plaquette unit was
\begin{equation}
H_\sigma = \frac{t'^2}{V}[const(x) + \vec{T}_A \cdot \vec{T}_B]+\frac{t'^2}{2V}
    \Big(\frac{1}{1+x}\Big)(T^y_A + T^y_B)
\end{equation}
The similar result for the perpendicular direction in the plane would be
\begin{equation}
H_\sigma = \frac{t'^2}{V}[const(x) + \vec{T}_A \cdot \vec{T}_B]-\frac{t'^2}{2V}
\Big(\frac{1}{1+x}\Big)(T^y_A + T^y_B)
\end{equation}
The minus sign for the single-plaquette terms in the second case is because the bond-ordering in the two perpendicular 
directions are the pseudospin in $+y$ and  $-y$ directions respectively(see Fig.\ref{fig:eigen}). Hence for the infinite square lattice, we get
\begin{equation}
H_{square} = \frac{t'^2}{V} \sum_{\la \alpha,\beta \ra}[const(x) + \vec{T}_\alpha \cdot \vec{T}_\beta]
\label{eq:TSquare}
\end{equation}
which is the antiferromagnetic Heisenberg Hamiltonian. Since, the square lattice is 
two-dimensional, its ground state will possess long range order. Notice that 
antiferromagnetic tendency of staggered pseudospin implies a ferromagnetic tendency 
for spontaneous currents in both ladder and the square lattice.

\subsection{Role of Symmetries}

Exploring symmetries can lead to a better understanding of the relation
between the form of microscopic model and that of the effective Hamiltonian.
Our starting fermion Hamiltonian (acting on a a two-plaquette unit) had
the following symmetries: a) time reversal
b) reflection symmetry (flipping the two plaquette upside down).
We have only considered models that maintain these symmetries.

\subsubsection{Consequences of generic symmetries}
\label{sec:symmetry_consequences}
                                                                                                                                                       
These symmetries imply specific symmetries in the pseudospin 
effective Hamiltonian \eqr{eq:Ham_spin}. (a)
The absence of a single-site term $\hat{P}^z$ follows from
the microscopic time-reversal symmetry,
under which the the {\it pseudocurrent} operator flips sign.
(b).  The absence of a single-site term $\hat{P}^x$ is due to the
transverse reflection symmetry of the two-plaquette unit,
under which the {\it charge-density} operator  flips sign.~\cite{FN-symmetries}

The tube and square lattices both have a 4-fold rotation symmetry, 
too,  under which the two bond-order states (pseudospin $+y$ and $-y$)
are equivalent, ergo the  $\hat{P}^y$ terms are absent.
Morever, for the most general one-particle spectrum that
a single plaquette could have
(keeping intact the degeneracy of momentum carrying states, i.e., $+\ti{1}$ and
$-\ti{1}$, but lacking particle-hole symmetry),
the effective Hamiltonian would still be of form Eq.~\eqr{eq:TLadder}, though
with different numerical coefficients.
On the other hand, the effective Hamiltonian 
for the ladder generically includes single-site $\hat{P}^y$ terms, 
since they are not ruled out by any symmetry.

\subsubsection{Role of lattice symmetry in ladder model}
\label{sec:lattice-symmetry-ladder}

As we just discussed, a four-fold lattice symmetry guarantees certain
pseudospin symmetries.
In our basic ladder Hamiltonian~\eqr{eq:Ham},
the single-plaquette terms had an ``accidental'' 
(non generic) four-fold rotational symmetry not guaranteed by the
ladder's symmetries.
If we generalize the ladder model so as to break the 
fourfold symmetry, 
what kinds of pseudospin asymmetries are generated?
                                                                                                                                                       
\begin{figure}[ht]
\includegraphics[width=0.8\linewidth]{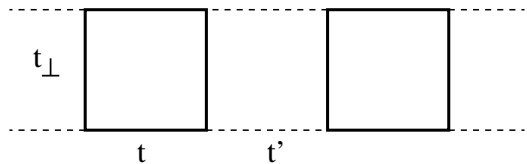}
\caption{A ladder without four-fold rotational symmetry around a plaquette.}
\label{fig:genladder}
\end{figure}
                                                                                                                                                
First we can make the transverse hopping $t_\perp$ within
a plaquette different from the longitudinal $t$
(Fig.~\ref{fig:genladder}).
Then the effective Hamiltonian
turns out to be
\newcommand{\Dt}{\Delta t}
\newcommand{\tbar}{\bar{t}}
  \begin{equation}\begin{split}
  H_{ladder} = &  const + \sum_\alpha
   \left(\frac{t'^2}{2V}\right) \vec{T}_\alpha \cdot \vec{T}_{\alpha +1} \\
  + \left(\frac{t'^2}{2V}\right)
      & \left(\frac{2\Dt}{(V-2\tbar+\Dt)(V-2\tbar+3\Dt)}\right)
                    T^y_\alpha  T^y_{\alpha +1} \\
      & +\left(\frac{t'^2}{2(V-2\tbar+3\Dt)}+2\Dt\right) T^y_\alpha
  \label{eq:tperp}
  \end{split}\end{equation}
where $\bar{t}\equiv (t+t_\perp)/2$ and $\Delta t\equiv t_\perp-t$.
The absence of four-fold symmetry of the hopping around the plaquette leads
to a first order field term $+2\Dt \, T^y$
and a second-order anisotropic exchange along the
$T_y$ (BDW) pseudospin direction for the ladder.
The underlying reason for this is that, so long as $t_\parallel=t_\perp$,
angular momentum is a good quantum number and in the $m=\pm 1$
single-fermion states, the sites from which fermions can hop differ
in phase by $\pm \pi/2$; the upshot is that {\it angular momentum 
is conserved} by degenerate hops, even though (in the ladder) it
is not conserved by excited-state hops, and thus the exchange terms
are isotropic.

\SAVE{I propose to hide this; it seems too obvious now.
``Since the tube and square lattice have a true four-fold symmetry,
it would not be natural to unsymmetrize the plaquette in those cases.''...agree}

Rather than spoil the ladder plaquette's four-fold symmetry in the hopping
terms, which couple to the pseudospin $T_y$ component,
we could do it by making {\it intra}-plaquette interactions
unequal, which produce single-pseudospin terms. 
Making transverse and longitudinal interaction different gives a
linear coupling at first-order in perturbation theory to
the bond-charge operator
$\propto (\VU_\parallel-\VU_\perp) T_y$;
as before, the time reversal and reflection
symmetries forbid linear  $T_x$ or $T_z$ terms.~\cite{FN-V2}

\subsection{Engineering spontaneous currents by fermion interactions (ladder)}
\label{sec:anisos-int}

Our study was motivated by the question : 
Can the microscopic models considered so far
exhibit current carrying states spontaneously in their ground 
state with genuine order in an Ising sense.
In the language of the pseudospin mapping [Eq.~\eqr{eq:oper_map},
we want the effective Hamiltonian Eq.~\eqr{eq:Ham_spin}
to give spin order along a particular axis.
But as we saw in Subsecs.~\ref{sec:Heff_tube} and \ref{sec:Heff_laddersq},
the natural form of \eqr{eq:Ham_spin} had a continuous XY symmetry 
(corresponding to the currents/charge operators) in the ladder case;
in the case of the tube or square lattice, \eqr{eq:Ham_spin} had the 
full three-component rotational symmetry of a Heisenberg magnet.
To stabilize any particular kind of order more than the others,
we must spoil these unwanted symmetries 
by anisotropic pseudospin terms
having the effect of an Ising-like anisotropy in
the desired direction.
In this subsection and the next, 
we continue the preceding one by
surveying various generalizations in the fermion Hamiltonian and
and the pseudospin terms they give rise to,
but now motivated by a
sort of engineering: rather than just solve for the ground
state of a given Hamiltonian, we frankly seek the Hamiltonian which 
gives the targeted state.

That need not require getting a uniaxial easy-$z$ anisotropy
(which -- see Subsec.~\ref{sec:guess-fermion} and \ref{sec:hops-general} --
is impossible from the fermion Hamiltonians we admit).  
Instead, we look for {\it two} perturbations that combine to 
{\it disfavor} the other two directions.
Now, since we have (pseudo)spin-1/2, there is no way to get any uniaxial
anisotropy from a single-spin term, but a trick is available: 
given we have antiferromagnetic order, a {\it uniform} external field 
perturbation creates an effective easy-plane for the directions normal
to the field; this was already used to get $xz$ anisotropy for the 
ladder case (from the linear $T^y$ term in \eqr{eq:TLadder}).
The other perturbation giving the needed hard-$x$ uniaxial anisotropy
must take the form of a pseudospin-pseudospin interaction, coming from
the {\it inter-plaquette} terms of the fermion Hamiltonian.


One natural extension of our model is to add an inter-plaquette 
nearest-neighbor interaction 
   \begin{equation}
    \label{eq:Vprime}
    H'= \VU' \big(n_A (1) n_B (0) + n_A (0) n_B (1)\big)
   \end{equation}
We have used the ladder numbering scheme in writing the above expression. 
Looking at Fig.\ref{fig:eigen}, we see how the term \eqr{eq:Vprime}
distinguishes the CDW sector from others, since the operators
like $n_A(1)$ depend on the CDW order; on the other hand, it
cannot distinguish different CCS or BDW states,
since they have equal fermion densities on all sites.
Using Eq.~\eqr{eq:oper_map}, we can easily convert the interaction
term to pseudospin language and indeed
   \begin{equation}
   \label{eq:Heff_Vprime}
    H'_{\sigma}  = \VU'\left(\frac{1}{2} - \frac{P_A^x P_B^x}{2}\right) =
       \VU'\left(\frac{1}{2} + \frac{T_A^x T_B^x}{2}\right)
    \end{equation}
We emphasize the above effective interaction \eqr{eq:Heff_Vprime}
is \emph{first} order 
in perturbation theory,  and not second order as for the hopping processes
earlier in Sec.~\ref{sec:effham}.
Eq.~\eqr{eq:Heff_Vprime} is an adjustment of the
$J_x$ pseudospin coupling in \eqr{eq:Ham_spin}
and thus favors CDW order, either a uniform pattern 
on each plaquette or an alternating one, depending on the sign and 
magnitude of $V'$. An inter-plaqutte second nearest-neighbor interaction
,  i.e.  $V'' (n_A (0) n_B (0) + n_A (1) n_B (1))$,
gives the same result as \eqr{eq:Heff_Vprime}
but with a flipped sign for the exchange term.

Thus, we see a route to favoring spontaneous currents for the ladder.
An infinitesimal attractive inter-plaqutte second nearest-neighbor interaction
 ($V' < 0$) or repulsive inter-plaquette second nearest-neighbor interaction
 ($V''>0$) will make $J_z > J_x$ and the ground state will have currents spontaneously.
For a comparison, we note that in generalized Hubbard models, 
attractive nearest neighbor interaction was argued to stabilize 
currents~\cite{stanescu_phillips}. 

For the tube and square lattice, inter-plaquette interactions can only
reduce the antiferromagnetic Heisenberg symmetry to a continuous XY symmetry
in currents/BDW plane and do not favor currents exclusively.
To do that, we must look to interplaquette hoppings instead.

\subsection{Engineering spontaneous currents by fermion hops (square lattice)}

An alternative extension of our model is to add additional inter-plaquette 
hoppings. As we will see, for the ladder and square lattice,
this favors bond (BDW) order by 
increasing the $J_y$ pseudospin exchange 
while decreasing $J_x$ and $J_z$ couplings in \eqr{eq:Ham_spin}. 
For the tube, inter-plaquette hopping to any distance
can never reduce the continuous Heisenberg symmetry due to 4-fold symmetry.


\subsubsection{Guessing the fermion term?}
\label{sec:guess-fermion}

A short-cut may allow us to quickly find fermion terms 
yielding a desired inter-plaquette pseudospin Hamiltonian form.
Let's extend the notion of ``pseudospin'' backwards to impute pseudospin
to the single-fermion states $|\pm 1\rangle$.  Indeed, we can 
just ignore the other orbitals, since only the
``degenerate'' hoppings  (explained at start of Sec.~\ref{sec:effham})
could give us a pseudospin interaction from second-order perturbation theory.
Then, we just substitute $\vec{T}_A \to \sum_{\sigma'\sigma} 
c\dagg_{\sigma'}(A) \vec{\tau}_{\sigma'\sigma}  c_\sigma(A)$, where
$\vec{\tau}_A$ means the usual Pauli matrices.  Thus, any coupling 
$\HH_{\rm  eff}^{T}$ between (components of) pseudospins
$\vec{T}_A$ and $\vec{T}_B$ gets transcribed to a four-fermion term
$\HH_{\rm  eff}^{c}$.
If we can regroup the four fermion operators so that
$\HH_{\rm  eff}^{c} \propto - {{\hat \HH}'}{}\dagg {\hat\HH}'$ where the 
operator ${\hat\HH}'$ hops a fermion from plaquette $B$ to $A$, 
then we could  take ${\hat\HH}'$ to be the inter-plaquette term 
reducing to $\HH_{\rm  eff}^{c}$ via second-order perturbation theory.

\SAVE{The energy denominator $\Delta E$, we saw in the first calculations,
is just a constant in the model that we set up.}

When applied to two plaquettes in a ladder or square lattice
(using the site labels of Fig.~\ref{fig:calcladder}), we get
   \be
     T^z_A T^z_B \to  
    i(c\dagg_{A,1}c_{A,0}- c\dagg_{A,0}c_{A,1}) \cdot
    i (c\dagg_{B,1}c_{B,0}- c\dagg_{B,0}c_{B,1})
   \eqend
and (for the bond-order component)
   \be
     T^y_A T^y_B \to  
     (c\dagg_{A,1}c_{A,0}+ c\dagg_{A,0}c_{A,1}) \cdot
     (c\dagg_{B,1}c_{B,0}+ c\dagg_{B,0}c_{B,1})
   \eqend
Thus, 
   \be
    \label{eq:T-to-fermions}
     T^z_A T^z_B + T^y_A T^y_B \to  
     (c\dagg_{A,1} c_{B,0}) (c_{A,0}c\dagg_{B,1})  +
    ( A \leftrightarrow B )
   \eqend
where there was only one of grouping such that the 
inter-plaquette hops connected nearest neighbors;
this is the term we found already.  
What about the term which would break the degeneracy 
between bond order and pseudocurrents:
    \be
    \label{eq:P-to-fermions}
     - T^z_A T^z_B + T^y_A T^y_B \to  
     ( c\dagg_{A,0} c_{B,0}) (c_{A,1} c\dagg_{B,1}) + 
    ( A \leftrightarrow B )
    \eqend
The grouped factors in Eq.~\eqr{eq:P-to-fermions} are 
diagonal (``$t'_{\sqrt 2}$'') fermion hops. 
Unfortunately, the sign of this term is necessarily
positive, so it always favors the $T_y$ (bond-order) 
direction.  

\subsubsection{Fermion hops in general}
\label{sec:hops-general}

A more comprehensive study of hops will be profitable for
the following reasons:
(i) As we are about to show, 
it reveals that the findings in \eqr{eq:T-to-fermions} and \eqr{eq:P-to-fermions}
are general for any nearest-neighbor hopping, so that is
not a route to the desired order.
(ii) Consequently,  further-neighbor
pseudospin interactions coming from long-distance fermion hops 
are our last hope to disfavor bond-order in the square 
lattice, and the general formula guides us to 
the correct interactions for this purpose
(iii) It is the root reason that the
ladder and square lattices' pseudospins needed to be staggered 
(at the beginning of Sec.~\ref{sec:effham}) but not the tube lattice's.

So imagine a perturbation Hamiltonian containing 
hops from {\it any} vertex of one strong plaquette to {\it any}
vertex of another (not necessarily the nearest neighbor).
First consider (still) the nearest-neighbor plaquettes, and
let $t'_1$, $t'_{\sqrt 2}$, $t'_2$, etc.
be (weak) hoppings to sites (in the other plaquette) at distances 
$1$, $\sqrt 2$, $2$ and so on, respectively.
Then the exchange part of the 
two-plaquette effective pseudospin hamiltonian is
\begin{eqnarray}
\label{eq:T-and-P}
H_\sigma & = & J_T \vec{T}_A \cdot \vec{T}_B + J_P \vec{P}_A  \cdot \vec{P}_B 
        \nonumber \\
& = & (J_T + J_P) P_A^y P_B^y - (J_T - J_P) (P_A^z P_B^z + P_A^x P_B^x) \nonumber \\
\end{eqnarray}
where 
\begin{subequations}
\label{eq:J_T-and-P}
\begin{eqnarray}
\label{eq:J_T_allhops}
   J_T &=& (t'_1 - 2 t'_{\sqrt 5}+t'_3)^2 / \VU > 0 , \\
   J_P &=& (t'_{\sqrt 2} - 2 t'_2+t'_{\sqrt{10}})^2 / \VU > 0
\label{eq:J_P_allhops}
\end{eqnarray}
\end{subequations}

\SAVE{SUMI: Do we need a figure showing the second-neighbor hops,
and also ${t''}_3$, ${t''}_{\sqrt 10}$, which figure in 
Sec.\ref{sec:uniform-2D}... Sumi thinks it's not necessary}

Which term does a given hopping contribute to?
First, remember we always need two different hoppings, 
with endpoints distinct; if they end at the same
site on one plaquette,  we could not couple to 
that plaquette's angular moment.  Then if we orient the 
two plaquettes such that the two hoppings don't cross, 
the exchange coupling relates staggered pseudospins or plain pseudospins
depending on whether clock sense on the two plaquettes are same
or alternating respectively. E.g., in Fig. \ref{fig:calcladder}, 
the non-crossed hoppings are connecting plaquettes with same clock 
sense, thus giving rise to staggered pseudospin exchange coupling,
i.e. $\vec{T}_A \cdot \vec{T}_B$. 

A related observation is that, in the ladder, there is 
a symmetry under mirror-flipping every second plaquette
around the long axis of the ladder, while
switching $t'_1 \leftrightarrow t'_{\sqrt 2}$ and
$t'_2 \leftrightarrow t'_{\sqrt 5}$;  Eq.~\eqr{eq:J_T-and-P}
shows this switches the $\vec{T}$ and $\vec{P}$ terms.

Inspecting \eqr{eq:T-and-P}, we see that so long as
we have only crossed or only uncrossed fermion hoppings, the result
is isotropic in the $(yz)$ pseudospin plane, so that bond
order and currents are degenerate.  However, if we
start from a mixture of crossed and uncrossed hoppings, 
the bond-order ($y$) exchange is always stronger than the
currents ($z$) exchange --- and is always antiferromagnetic.

\subsubsection{Spontaneous currents via anisotropic frustration}
\label{sec:aniso-frustration}

Given this last fact, is it possible at all to obtain a
pseudospin anisotropy favoring $T_z$ and hence current order
over the whole lattice, by coupling more distant units?
This is possible, in principle, through {\it anisotropic frustration}. 
(It is assumed interactions have somehow already 
disfavored charge ordering, as discussed in subsubsection 
\ref{sec:anisos-int}.)

Assume the dominant nearest-neighbor hopping is purely
$t'_1$, as in our original and simplest model.
The pseudospin exchange has a continuous symmetry in
the $yz$ spin plane, leading to  antiferromagnetic order 
degenerately in any mixture of those components.
Now imagine (say) a second-neighbor pseudospin exchange
due to mixed kinds of hoppings;  by the above arguments,
$J_{2y}$ is necessarily antiferromagnetic, and $J_{2y}> |J_{2z}|$.
But unlike the nearest-neighbor exchange, the enhanced second-neighbor
$J_{2y}$ term {\it disfavors} bond-order state (being of the
wrong sign).

\SAVE{The key fact is that the most important distant hops 
are crossed with respect to each other, in the opposite sense 
from the nearest-neighbor hops.}  

A second-neighbor exchange is allowed on the ladder, 
using (say) the hops ${t''}_3$ and ${t''}_{\sqrt{10}}$
connecting two plaquettes related by a $[4,0]$ vector.
On the square lattice, the second nearest neighbor has
a displacement [2,2] and this exchange turns out to be 
symmetry-forbidden.
But the square lattice can have the same $[4,0]$ 
inter-plaquette hops as on the ladder, and these 
finally give our goal: we can favor spontaneous
order in the square lattice, albeit with a rather
baroque Hamiltonian.

\SAVE{There will be symmetry-related hops, e.g. ${t''}_{\sqrt 5}$),
related by symmetry about the $[1,1]$ diagonal line 
connecting  the plaquette centers.  It turns out
two terms combine with a relative phase shift of
$\exp(i 2 \frac{\pi}{2}) = -1$ to cancel out each other.
We also have $[1,1]$ hops, but as their connecting points
are $\pi$ apart around the plaquette, these can't possibly
distinguish the sense of currents, so those terms are
zero too.}



\section{Generalization to Spinfull Models}
\label{sec:spinfull}

It is natural to ask if we can extend our results to models with spin,
In case they can be applied to a real electronic system, 
and also to make some contact with other works on spontaneous current
models.
There are two quite different ways to imagine this. 
First, as worked out in Subsec.~\ref{sec:addspin}, we can simply 
include an additional spin degree of freedom in the Hamiltonians 
considered above.
Alternately, as worked out in Subsec.~\ref{sec:nersesyan}
we can exactly map a site degree of freedom in 
one of our spinless models to the spin degree of freedom in
a model with half as many of lattice sites 
(thus keeping constant the the total degrees of freedom.)

\subsection{Adding Spin Degree of Freedom}
\label{sec:addspin}

For this extension of our model, we simply add spin indices in all the terms
of \eqr{eq:Ham} while conserving the spin, and rerun the calculations of 
Secs.~\ref{sec:setup} and \ref{sec:effham}.  

To make the spinfull calculation analogous to what we did, the filling 
should now be $3/8$.
Of the three fermions per plaquette, 
the first two fill angular momentum zero 
($|\tilde{0} \uparrow \rangle$ and $|\tilde{0} \downarrow \rangle$).
The third fermion goes in the degenerate current carrying state, 
$|\pm \tilde{1} \sigma \ra$.
The extended pseudospin representing each plaquette 
is now the direct product 
of the same pseudospin degree of freedom, and a
real spin $\vec{S}_\alpha$ ~\cite{FN-spinfull-half}. 
Also, the interaction term in Eq. \eqr{eq:Ham-V} has to be augmented
to include an onsite interaction term ($U$) equal in strength 
to the offsite interaction terms ($V$) so that the multi-fermion
eigenstates still remain direct products of single-fermion
orbitals.

Here are the results for each case:

Spinfull Tube : 
\begin{eqnarray}
\label{eq:spinfull_tube}
H_{eff} &=& \left(\frac{2 t'^2}{V}\right) \sum_{\alpha}  \Big[ (\vec{P}_\alpha \cdot \vec{P}_{\alpha + 1})  +
 (\vec{S}_\alpha \cdot \vec{S}_{\alpha + 1})  \nonumber \\ 
& & + 4 (\vec{P}_\alpha \cdot \vec{P}_{\alpha + 1}) (\vec{S}_\alpha \cdot \vec{S}_{\alpha + 1}) \Big]
\end{eqnarray}

Spinfull Ladder : 
\begin{eqnarray}
H_{eff} &=& \left(\frac{t'^2}{2V}\right) \sum_{\alpha} \Big[ \left(\frac{1}{2}\right) \vec{T}_\alpha \cdot \vec{T}_{\alpha + 1} + \\
& & (\vec{T}_\alpha \cdot \vec{T}_{\alpha + 1})(\vec{S}_\alpha \cdot \vec{S}_{\alpha + 1}) - \left(\frac{1}{1+x}\right) T^y_\alpha \Big] \nonumber 
\end{eqnarray}

Spinfull Square Lattice : 
\begin{eqnarray}
\label{eq:ham-spinfull-square}
H_{eff} &=& \left(\frac{t'^2}{2V}\right) \sum_{\la \alpha,\beta \ra}  
    \Big[ \left(\frac{1}{2}\right) \vec{T}_\alpha \cdot \vec{T}_\beta + \nonumber \\
      & & \qquad (\vec{T}_\alpha \cdot 
\vec{T}_\beta)(\vec{S}_\alpha \cdot \vec{S}_\beta) \Big]
\end{eqnarray}
Thus the effective Hamiltonians have a form like the Kugel-Khomskii 
Hamiltonian~\cite{kugel-khomskii} for cubic titanates describing spin and orbital 
superexchange interactions between $d^1$ ions having threefold degenerate 
$t_{2g}$ orbitals. 

The result \eqr{eq:spinfull_tube} for the tube case is proportional
(modulo a constant) to
$(\half + 2 \vec{P}_\alpha \cdot \vec{P}_{\alpha + 1})$
$(\half + 2 \vec{S}_\alpha \cdot \vec{S}_{\alpha + 1} )$
which is the $SU(4)$ symmetric 
Kugel-Khomskii model~\cite{kugel-khomskii-SU4}.
For the tube, the interaction terms are just ``degenerate" hops of 
Sec. \ref{sec:effham}; they conserve spin as well as pseudospin. Actually,
they conserve a combined flavor which includes both the spin and the pseudospin.
Hence, the effective Hamiltonian possesses an $SU(4)$ symmetry in 
which there is no distinction between the four combined flavors the 
hopping fermion might carry. The interesting behavior of such $SU(4)$
chains is discussed in Ref.~\onlinecite{kugel-khomskii-SU4};
in terms of the original fermions, it obviously corresponds to a
high degeneracy between many kinds of order.

For the ladder and square lattice cases, the
``degenerate" hops do not conserve the combined flavor thereby reducing
the $SU(4)$ symmetry to only $SU(2) \times SU(2)$ for the exchange terms.
What kind of order do these lattices have?  Notice that the
spin-pseudospin cross-terms tend to favor ferromagnetic order
in one kind and antiferromagnetic order in the other.
Since we also have a pseudospin antiferromagnetic exchange 
but no real spin exchange, the expected order is always
{\it ferromagnetic} for the real spins~\cite{FN-spinfull-analysis}.
Then the pseudospin order
is the same as in a spinless model; in effect, the system 
spontaneously becomes spinless by polarizing in one spin flavor.

We compare our results to that of \cite{Yao_Kivelson}, in which a 
Hubbard model with a similar pattern of strong and weak plaquettes
($t$ and $t'$) with just an onsite interaction term was studied 
on a square lattice. 
Yao {\it et al} found a host of different phases for different values of the onsite
interaction including a Fermi Liquid, a $d$-BEC, a $d$-CDW, a $d$-BCS,
a spin-$1/2$ antiferromagnet, a spin-$3/2$ antiferromagnet and an ``orbital nematic"
phase at $3/8$ filling, which is equivalent to one doped hole, with respect
to half filling, on each plaquette
 (``$Q_h=1$'' or ``$x=1/4$" in their notation). 
In their model, it is only in the parameter regime
$U_c \simeq 4.6 t < U < U_t \simeq 18.6 t$ that their
single-plaquette states are filled like ours and 
admit the possibility for currents~\cite{FN-Yao-outside}
Then, the single-plaquette states are characterised by spin-$1/2$ 
as well as a pseudospin-$1/2$ (called chirality $\tau_z$ 
by Ref.~\onlinecite{Yao_Kivelson},
and having ``$p_x \pm i p_y$" symmetry i.e.
our angular momentum $|\pm \ti{1}\ra$.) 

\SAVE{Change to SAVE:
Specifically, below $U_c$, their doped holes (or equivalently
their fermions) bind together to form effective hard-core bosons 
and give rise to the $d$-BEC, $d$-CDW and $d$-BCS phases;
in other words, occupation $0$ and $2$ is more stable than 1.
Since the doped holes bind together, there is no possibility
for orbital currents. 
Above $U > U_t \simeq 18.6$ the doped hole becomes a spin-$3/2$ 
object with $s$ wave symmetry. Thus,
the only parameter regime that admits orbital currents in their setup 
is $U_c < U <U_t$.}


They do not discuss the possibilities of orbital currents
explicity in the above mentioned regime, but implicit in their result are the 
anisotropies in the
pseudospin exchange terms which is interesting (See Eq. (4) and (6) of \cite{Yao_Kivelson}). 
They state that, in the regime $U_c < U <U_n \simeq 7.3$, the system becomes a spin-$1/2$
antiferromagnet with electron nematic order (same as our BDW), while in $U_n < U <U_t$,
there is no nematic ordering. Perhaps, there are spontaneous currents in this regime.
However, they do not discuss the origin of the pseudospin anisotropies. It is all the
more perplexing to us, given our experience that one set of ``degenerate" hops taking
part in the second-order perturbation theory can only give rise to isotropic exchange.


\subsection{The Nersesyan Map}
\label{sec:nersesyan}

As first proposed by Nersesyan~\cite{nersesyan}, a spinfull
model on any lattice can be mapped to a spinless model 
on a doubled version of that lattice (its direct product by
$\{1,2\}$.)
Each pair of sites in the spinless model
represents respectively the spin-up and spin-down occupation.
Thus, a spinless ladder maps to a spinfull chain (or vice versa), 
such that the leg index maps to the spin index.
(We shall call this a ``rung spin'' to distinguish it from
the real spin of Sec.~\ref{sec:addspin} and the pseudospin of
all the earlier sections.)
Hamiltonian terms acting on a single rung of the ladder will be mapped 
to single-site terms on the chain,
while terms along the ladder's leg map to terms along the chain's leg. 
We exhibit examples of the map in both directions.

In fact, since our plaquette is built from two rungs,
each plaquette pseudospin operator $\vec{P}_{i}$ corresponds
to two neighboring rung spin 
operators as shown below.  
Consider the fermion basis states on each rung $j$ that 
have nonzero pseudocurrents, namely 
   \begin{subequations}
   \label{eq:our-rung-spins}
   \begin{eqnarray}
     |\phi_{j+}\ra  &\equiv&  \Big(|j,1\ra + i |j,2\ra\Big)/\sqrt{2}, \\
     |\phi_{j-}\ra  &\equiv&  \Big(|j,1\ra - i |j,2\ra\Big)/\sqrt{2}.
   \end{eqnarray}
   \end{subequations}
The $\pm$ label  in \eqr{eq:our-rung-spins}
is a rung pseudospin index defining the $z$-axes for rung pseudospin $\vec{S'}_j$
\emph{aligned} with that of the plaquette pseudospin such that 
\begin{subequations}
   \label{eq:rung-pseudo}
   \begin{eqnarray}
   	\vec{S'}_{2j} & = & (P^x_j, P^y_j, P^z_j) \\
	\vec{S'}_{2j+1} & = & (-P^x_j, P^yj, -P^z_j)
   \end{eqnarray}
   \end{subequations}
On the other hand, Nersesyan's rung spins (we call $\vec{S}$ keeping
in mind the difference from the previous section) are  related to
our rung spins $\vec{S'}$ via (See Eq. 3-5 of \cite{nersesyan})
   \begin{subequations}
   \label{eq:rotate-nersesyan}
   \begin{eqnarray}
      S^x_j &=& (-1)^j S'^y_j  \\
      S^y_j &=& (-1)^j S'^z_j  \\
      S^z_j &=& (-1)^j S'^x_j
   \end{eqnarray}
   \end{subequations}
Notice that the definition of Nersesyan's rung spins is staggered 
compared to ours.
Nersesyan used a ``canted" rung spin basis to make manifest the staggered nature
of charge/spin densities in CDWs and Spin Density Waves(SDW) for a spin-$1/2$ Hubbard chain.
 We will write rung spins using the Nersesyan basis from now on.
Using \eqr{eq:rotate-nersesyan}, we can re-use the results of
Sec.~\ref{sec:effham} to figure out what a given fermion perturbation
projects to in terms of Nersesyan pseudospins.

\SAVE{CLH has not checked the above in detail as of 11/4/08, but trusts it.}
We shall return to apply the Nersesyan map in Sec.~\ref{sec:uniform-nersesyan}.

\subsubsection{Spinless ladder to spinfull chain}

In particular, our basic Hamiltonian for the spinless ladder 
[see Eq.~(\ref{eq:Ham})] gets mapped to a chain of alternating
strong and weak bonds
\newcommand{\spf}{^{\rm sf}}
(``sf'' here  distinguishes the spinfull model parameters):
  \begin{eqnarray}
       \label{eq:Spinmap}
       \HH   &\equiv & - \sum _{i, \sigma} 
                   t\spf_{i,i+1} [c\dagg_{i\sigma}c_{i+1,\sigma} + h.c.] 
        \nonumber \\
         &+& U\spf \sum_i \hat n_{i\uparrow} n_{i\downarrow} 
        -h\spf_x \sum _i  S_i^x 
        + \sum_i  V\spf_{i,i+1} \hat n_i  n _{i+1} \nonumber 
        \label{eq:SM-tprime}
  \end{eqnarray}
with $h\spf_x=t$, $U\spf=\VU$; we get $t\spf_{i,i+1} = t$ or $t'$, 
and $V\spf_{i,i+1} = \VU$ or 0, respectively for strong or weak bonds
$(i,i+1)$.
Thus our spinfull chain includes the usual hopping and interaction 
a nearest neighbor interaction and a field term along $x$ in spin-space  
or a single-site spin flip term. This kind of spin mapping will map 
spontaneous current states to 
spin current states with canted site-spin expectations, 
CDW to SDW, and BDW to an equivalent BDW/paramagnet 
as one may readily verify. The advantage of this kind of mapping is that
we may carry known results from 
spinfull models to spinless models  or vice versa.

\subsubsection{Spinfull ladder to spinless tube}

Tsuchiizu and Furusaki~\cite{furusaki} presented a spinfull ladder model
which had a spontaneous current phase.   Their Hamiltonian included the 
following kinds of parameters 
Two hoppings (longitudinal ${t\spf}_\pa$ and rung ${t\spf}_\perp$); 
an on-site Hubbard repulsion $U\spf$;  three neighbor repulsions
(ladder $V\spf_\pa$, rung $V\spf_\perp$, and second-neighbor ${V\spf}'$);
a spin exchange $J\spf_\perp$, acting across a rung; and, last but
not least,  a correlated hopping or ring exchange $t\spf_{\rm pair}$, 
which takes two electrons from one site to the other site
on the same rung.

Under the ``Nersesyan map'', their model corresponds to a kind of
``tube'' model with spinless fermions, but of course lacking the 
4-fold symmetry of our tube model.
Their repulsions $V\spf_\perp$ and $U\spf$, as well as the
$zz$ term of the exchange interaction, map to various on-plaquette
repulsions like out $\VU$; 
while their $V\spf_\perp$ and ${V\spf}'$ map to combinations
of inter-plaquette interactions in the spinless model, including 
(but not limited to) our $V'$. 
Their $t\spf_\pa$ is the same as our inter-plaquette hopping $t'$,
but their rung hopping $t\spf_\perp$ corresponds to hopping on
only two edges of the plaquette in the spinless model.
(An  $x$-oriented magnetic field -- not included in their model --
would have mapped to spinless hopping on the plaquette's other edges.)
Finally, their spin-exchange $J\spf_\perp$ and correlated hop $t\spf_{\rm pair}$
both map to correlated (two-fermion) hops, like no term in our models:
the $xy$ component terms of $J\spf_\perp$ would correspond 
to two diagonal spinless fermions on 
one plaquette moving to the opposite diagonal, while $t\spf_{\rm pair}$
takes two adjacent fermions on one side of a plaquette and moves 
them to the two opposite sites.

The strong coupling approach of Ref.~\cite{furusaki} corresponds
exactly to our separation of ``weak'' and ``strong'' plaquettes.
Their state with spontaneous (pseudo)currents is somewhat
disappointing, as the return current depends on their pair hopping term.
(Of course, in this limit that all inter-rung couplings go to zero,
it would not be possible to have a genuine current involving two
sites, since conservation requires a loop for it to circulate around.)

Sch\"ollwock {\it et al}~\cite{schollwock-ladder} exhibited a 
simpler spinfull ladder 
which was shown numerically to have a spontaneous-current phase.
They have just one hopping parameter $t\spf_\perp = t\spf_\pa = t\spf$ 
(thus they could not access the  strong-coupling limit of small $t\spf_\pa$).
They also have Hubbard $U$, exchange $J\spf_\perp$, and repulsion $V\spf_\perp$.
In the special case that $J\spf/4 = U\spf - V\spf_\perp$, this maps
to a tube model with fourfold symmetry 
(this is {\it not} the same special point 
$J\spf/4 = U\spf + V\spf_\perp$, 
which had $SO(5)$ symmetry~\cite{schollwock-ladder}).
The mapped spinless model would also have 
the same correlated hop of diagonal pairs as Ref.~\onlinecite{furusaki}.

\section{Towards Uniform Spinless Models}
\label{sec:uniform}

Our hope was that if we can find a strong-coupling model that
has spontaneous currents, perhaps it can be adiabatically continued
to a translationally invariant model that does not distinguish
the Hamiltonian terms on weak and strong plaquettes.
There are two preconditions to even consider this:
\begin{itemize}
\item[](i) The symmetry pattern of order should be consistent with
a uniform order. Below, in subsec.~\ref{sec:I_weak}, we verify this
for our models.  
\item[](ii) The ``weak'' and ``strong'' terms in the
Hamiltonian should have the same form, differing only by
the size of the coefficient.
That is easy enough to manage, even if we adhere to the 
somewhat unnatural interaction term \eqr{eq:Ham-V},
with the nearest- and second-neighbor strengths made
equal for convenience.  If that equality is carried over 
to the inter-plaquette terms, it is actually beneficial
since it cancels a term favoring CDW ordering
[see Eq. \eqr{eq:Heff_Vprime}]
\end{itemize}

There is one further challenge: having conjectured a 
Hamiltonian favoring current ordering, how could we
verify that?  We need some family of variational
wavefunctions that would (ideally) be definable for
all the interpolating Hamiltonians from strong coupling
to uniform, and where the variational parameters allow
any value of the spontaneous-current order parameter.
(The best model of such a calculation is the
recent work of Ref.~\onlinecite{weber-giamarchi}.)
In the case of the spinless ladder models, such a 
correspondence was already worked out by Nersesyan~\cite{nersesyan} 
and is elaborated below in Sec.~\ref{sec:uniform-nersesyan}.

\subsection{Inter-plaquette Spontaneous currents}
\label{sec:I_weak}
                                                                                                                                                             
Previously, we saw that at least for the ladder, the ground state could exhibit spontaneous currents by producing 
anisotropy along that sector. Such a symmetry-broken state has spontaneous current expectations not only on the 
strong plaquettes, but also on the weak bonds connecting them. Depending on the symmetry of this pattern, it may 
be possible to adiabatically connect to a state in which the strong/weak distinction goes away and the current 
expectations have equal strength on all bonds.
                                                                                                                                                             
{\bf Current expectations on Weak Bonds:}
For the weak bonds, the required pseudo-current operator is 
$\hat{I}_{weak} = i[c\dagg_{Ai} c_{Bj} - c\dagg_{Bj} c_{Ai}]$ 
where $i-j$ is a weak bond. We again use the canonical 
transformation recipe, \eqr{eq:optran} from the appendix, 
but now for the purpose of deriving an expectation rather
than a Hamiltonian term.
The pseudo-current operator for the weak bonds projects, in 
pseudospin language, to
\begin{align}
\label{eq:weak-current}
\hat{I}_{weak} = & - \frac{t'}{2(2t + \VU)} 
\Big(|2+;2+\rangle \langle2+;2+| -  \nonumber \\
     & \qquad |2-;2-\rangle \langle2-;2-|\Big) \nonumber \\
 = &  - \frac{t'}{2(2t + \VU)} (P^z_A + P^z_B) \nonumber\\
 = &  - \frac{t'}{2(2t + \VU)} (T^z_A - T^z_B)
\end{align}
Thus, in the ladder or square lattice where $P_z$ tends to be
uniform, i.e. $T_z$ is staggered, we see the (pseudo)currents 
in weak bonds are proportional to those in the adjoining strong bonds: 
both kinds of current must have the same degree of order.
However, in the tube lattice, the weak 
bond currents are zero.
That must be true to all orders on grounds of symmetry.
(Reflection in the plane of a strong plaquette reverses 
the sense of currents on weak plaquettes but not on
strong ones.)

\begin{figure}[ht]
\includegraphics[width=0.5\linewidth]{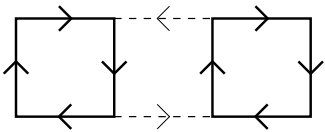}
\caption{The orbital current pattern in $|2+;2+\rangle$ state.}
\label{fig:ddwlike}
\end{figure}

Since the pseudospin operator for the strong bond is just $+\hat{P}^z/2$ 
[Eq.~\ref{eq:oper_map}], 
the weak-bond and strong-bond currents have {\it opposite} directions
(Fig. \ref{fig:ddwlike}), which is the same pattern as the 
$d$-density wave state~\cite{ddw}.
Thus, this pattern is consistent with 
a ddw-like state if we could analytically continue our
model to a uniform one.

\subsection{Uniform model}
\label{sec:uniform-nersesyan}

\SAVE{There were actually two regimes, 
depending on the ratio $t/\VU$.
If $t/\VU \gg 1$, we could ignore the excited-state hops, which 
give a $T_y$ effective field...}

This is \emph{not} the same as our pseudospin map of Sec.~\ref{sec:pseudospin}.

Can one of our strong-coupling models be connected to a uniform one, 
in which the distinction of strong and weak bond vanishes? 
We found the Nersesyan map [Sec.~\ref{sec:nersesyan}]
was the key to guessing a Hamiltonian that 
has a spontaneous-current ground state.  The spinless model's
pseudocurrent order maps (as we show shortly) to a certain spin order
in the spinful model.
One need only invent a \emph{uniform} spin Hamiltonian 
giving the desired order, 
and such that it maps to a plausible Hamiltonian of the spinless fermions.
The only limitation is that the resulting spinless model is built from
site pairs: it lends itself to ladders, or bilayers, but not to (say)
the fourfold symmetric square lattice.

The Nersesyan approach is an existence proof that if we get spontaneous
current order in a simple, local, strong-coupling model, the adiabatic
extension can work all the way to the uniform case.  If we could
find an analogous starting point on the square lattice, that result
would make the corresponding extension more plausible; unfortunately
(Sec.~\ref{sec:uniform-2D}) we could not find such a starting point.

Concerning the certain spin order mentioned above, consider
a strong plaquette in our ladder with angular momentum $+1$
which has spontaneous pseudocurrents on its rungs and 
leg bonds. It maps to a pair of sites on the spinfull chain
with Nersesyan spin expectations 
   \be
   \label{eq:nersesyan-spin-order}
       \langle \vec{S_j} \rangle \propto (1,(-1)^j,0)
   \eqend
which corresponds correctly to a non-zero $\la P^z_j \ra  =1$
(See Eq. \eqr{eq:rung-pseudo} and Eq. \eqr{eq:rotate-nersesyan}).
\SAVE{Sumi says 11/08:  Yes, these signs are right for ang. momentum $+1$
Yes, the $x$ and $y$ components are exactly equal to each other.}
The uniform  $S_x$ component is not surprising,
as every spinless ladder we consider includes
a rung hopping which becomes an $x$ transverse field
under Nersesyan's map.

The pseudocurrent along a rung, $I^\perp_i$, maps to $S_y$,
corresponding to the alternating $y$ component.
What about the pseudocurrents along the legs?
Let's define a difference between the two sides,
   \begin{equation}
     I^\pa_i \equiv I_{(i,1)\to (i+1,1)}- I_{(i,2)\to (i+1,2)} .
   \end{equation}
Of course, in the expected state, opposite sides have opposite
currents, so this also has a nonzero expectation.
Evidently, this simply maps to the $z$ component of {\it spin current} from
$i \to i+1$,   $I^S_{i\to i+1}$.
Indeed, in spin models with an isotropic Hamiltonian (and also here),
spin current goes with the twist between noncollinear spins,
$\vec{I}_{i\to j} \propto \SS_i \times \SS_j$
\SAVE{This sign is correct}
and indeed, the staggering of rung-spin directions
in \eqr{eq:nersesyan-spin-order}
does give the requisite nonzero (and alternating) 
$(\SS_i\times \SS_{i+1})_z$ component.

Now we see a simple route to a {\it uniform} model having spontaneous
currents: simply find a uniform Hamiltonian for a spin-$1/2$ chain, having
a ground state with the above staggered spin canting.  The simplest 
realization (from the spin chain viewpoint)
would be an antiferromagnetic chain with (i) an $S_x$ field, 
so the spins will cant transverse to it -- recall this is 
a simple transverse hopping in the spinless model;
plus (ii) a small anistropy in the antiferromagnetic exchange, 
such that the $S_y$ axis is easier than $S_z$; this
ensures the canting happens in the $y$ direction.
The problem with following this route literally is ingredient (ii):
$S_x$-$S_x$ or $S_y$-$S_y$ spin couplings correspond, in the 
spinless model, to two-fermion pair correlated hops in a plaquette, 
which we did not want to include. Conversely, this nicely
illustrates why correlated hopping is conducive to the 
existence of spontaneous currents \cite{nayak} !

The scenario of the previous paragraph can be achieved, instead,
with the following adjustment of the Hamiltonian: in place
of a spin chain, take a {\it Hubbard} chain with 
a transverse $x$ field $h\spf_x$, as above, plus a small 
{\it ferromagnetic} coupling $J\spf_z$ of neighboring $S^z$
components.
A strong Hubbard on-site repulsion $U\spf$ provides the 
effective antiferromagnet exchange at order $(t\spf)^2/U\spf$, 
in the standard fashion. Via the Nersesyan map (see Eq. \eqr{eq:rung-pseudo}
and \eqr{eq:rotate-nersesyan}), we realise that: 1) the transverse field suppressing
spin ordering in $x$ direction maps to a term suppressing
of BDW in the spinless ladder, 2) the small ferromagnetic $z$ coupling
suppressing spin ordering in $z$ direction (by reducing the 
strength of $z$ antiferromagnetic exchange) maps to a term suppressing
CDW in the spinless ladder, and finally 3) due to the aforesaid suppresions,
stabilization of spin order in the $y$ component maps to orbital currents
for the spinless ladder.

In the spinless language,
the Hubbard $U\spf$ and $-J\spf_z S^z_i S^z_{i+1}$ interactions just map,
respectively, to nearest-neighbor interactions along the rungs and legs 
(repulsive $V_\perp$ and attractive $V_\pa$).
The Hubbard hopping $t\spf$ and the transverse field $h\spf_x$ 
just map, respectively, to hoppings along the rungs and legs
(our $t_\perp$ and $t_\pa$).
The half filling we adopted for the spinless model corresponds
to half filling in the Hubbard model.
Unlike all spinless models we previously mentioned, 
this model is a uniform ladder with {\it no weak and strong plaquettes}.

The above paragraph is essentially a rediscovery of Nersesyan's ladder model, 
Eq. (1) of \cite{nersesyan}.  
\SAVE{Added; is this right?.. yes}
In particular, our key ingredient --
making $V_\parallel$ attractive while $V_\perp$  repulsive and 
zero $V_{\sqrt{2}}$ -- is
essentially the same as Nersesyan's recipe, which is that
$V_{\sqrt 2} - V_\parallel > 0$.
It is interesting to note these chains have fractionalized excitations, 
domain walls carrying 1/2 fermion charge~\cite{narozhny},
corresponding to spinons in the spin model.

\subsubsection{Extension to the square lattice?}
\label{sec:uniform-2D}

The same mapping can be used in one dimension higher, to build
a spinless spontaneous-current state on a bilayer from a Hubbard 
model on a square lattice~\cite{kolezhuk} (or any bipartite lattice).
A {\it spinfull} model with plausible
interactions on that same lattice was known earlier~\cite{capponi-bilayer}
that exhibits spontaneous currents.  
That model's Hamiltonian is similar to the Schollw\"ock {\it et al}
ladder~\cite{schollwock-ladder} we described above, in one dimension
higher, and is similarly engineered to have an SO(5) symmetric point.
(A minor difference is their ``rung'' (interlayer) hopping $t\spf_\perp$ 
may differ from the in-layer hopping $t\spf_\pa$.)  
All their interaction terms -- $U\spf$, $V\spf_\perp$, and (isotropic) 
exchange $J\spf_\perp$ -- act only on rungs.  Thus, the main qualitative
difference our model  differs (apart from spin, of course) is our
in-layer repulsion term $V_\pa$.  The (conjectured) order in our 
bilayer model is the same alternating pattern of currents as in
their model (Fig.~1 of Ref.~\onlinecite{capponi-bilayer}).

Unfortunately, this does not work for the plain square lattice, for
two reasons \cite{sun}. Firstly, the best we could manage for a Hamiltonian
(Sec.~\ref{sec:aniso-frustration}) depended on ${t''}_{3}$ and ${t''}_{\sqrt 10}$
hoppings to a third neighbor plaquette, while the much shorter
hoppings entering~\eqr{eq:J_P_allhops}
must be negligible.
Not only are those absurdly distant neighbors to have a meaningful hopping: 
once we make the lattice uniform, we must also include 
${t'}_{3}$ and ${t'}_{\sqrt 10}$ of the same separations as
${t''}_{3}$ and ${t''}_{\sqrt 10}$ 
but coupling {\it nearest} neighbor plaquettes and 
creating the undesired un-frustrated anisotropy \eqr{eq:J_P_allhops}.
Secondly, we cannot be guided by Nersesyan's map, as it demands that sites
come in {\it pairs}.  

Although, it seems difficult to obtain currents order in a uniform square latttice,
the terms which worked in the ladder would still be
effective in a rectangular lattice.
That suggests one possibility to obtain spontaneous currents 
as a symmetry breaking.
Notice that a {\it uniform} bond order, 
i.e. the state currently called ``electron nematic''~\cite{electron-nematic}
reduces the system  to rectangular symmetry, making $t_\parallel \neq t_\perp$
heuristically (See Sec.~\ref{sec:lattice-symmetry-ladder}).
Thus, spontaneous currents could be parasitic on electron nematic
order.  This is a not a linear coupling of the two order parameters; it would be 
a second Ising-like transition, to a state breaking $Z_2 \times Z_2$ symmetry.

\SAVE{Convert to SAVE.  Does this effect does NOT survive square symmetry?
We must ask what is the pattern of the natural bond order, which would
be staggered with respect to an external uniform nematic order. 
We assume strong bonds do not like to be facing each other across 
a plaquette. If they nevertheless do form a dimer covering, the only
choice on the square lattice is the alternating pattern which is
the state of maximum tilt in the height representation.   
(That was well studied circa 1989 from the quantum dimer models inspired by
quantum spins.)  Then a uniform bond order would indeed be staggered
relative to this order.}

\section{Discussion}
\label{sec:discussion}

\SAVE{I would talk about future work, but we didn't really have anything to say.}
The central aim of this paper was to investigate the possibilities of a
lattice model \emph{manifestly} displaying spontaneous currents in its
ground state. Among our models,
containing standard hopping and interaction terms,
it was fairly difficult to stabilize only
the spontaneous-current state.  Frequently, there was a
remnant continuous symmetry leading to an arbitrary mixing with one of the
competing orders; and most perturbations which could break
that degeneracy tended to favor the competing order.
Something similar also happens in some spinfull models meant to
address the possibility of spontaneous currents in a realistic
system, e.g. the relation~\cite{kim} of ``d-density wave''
current order and ``electron-nematic'' order (related to our
``bond order'').
                                                                                                                                                             
We did show that rather contrived and unappealing fermion interaction
or hopping terms
could be used to stabilize currents, but it seems highly unlikely
that those kinds of  processes are at work in real materials.
We suggest our finding may be related to the rarity of solids in nature
having spontaneous currents in their ground state \cite{sun}.

\subsection{Summary}




The central results of this paper are as follows. 
We emphasize first that our microscopic Hamiltonian was limited to 
(mainly spinless) models  with interactions and one-fermion hopping terms. 
We did not explore the possibilities of correlated hopping, which were
already known to be conducive to the ``d-density-wave'' 
current order~\cite{nayak, furusaki}.
Ultimately such terms come microscopically from higher-order processes 
in fermion hops; thus, within our picture, related terms might be
accessed by expanding our canonical transformation (Sec.~\ref{sec:effham}
and Appendix~\ref{app:canonical}) to higher orders, producing
effective interactions with higher powers of pseudospin.

We showed that by tuning the parameters and the underlying geometry
of a toy spinless Hamiltonian, we can make a system
acquire spontaneous currents, charge or bond order. The crucial ingredient
of our analysis was the pseudospin mapping (Sec \ref{sec:pseudospin})
and degenerate second-order perturbation theory (Appendix \ref{app:canonical}).
We saw that bond ordering is naturally
disfavored in ladders (Sec \ref{sec:Heff_laddersq}); 
while, for tube and square lattice, the ground state can acquire possibly coexisting charge,
bond or current order in a symmetry breaking fashion (Sec \ref{sec:Heff_tube} and \ref{sec:Heff_laddersq}). 
Since, the tube is quasi one dimensional, the correlation will actually be power laws; but for the square lattice, we will have  true long range order. 
Furthermore, the pattern of currents
corresponding to the spontaneous currents carrying ground state is same as the 
d-density wave state (Sec. \ref{sec:I_weak}).
The Nersesyan map provided a way to extend our strong-coupling result to that of an uniform case 
for the ladder (Sec \ref{sec:uniform-nersesyan}).

\subsection{Relation to three-band model and other real systems}

The current contact of our topic with real systems is in the
three-band model of cuprates~\cite{varma,simon-varma}.
A recent paper~\cite{new-giamarchi}
argued (by mapping to a 2-channel Luttinger liquid and
then analytic perturbation) that a ladder version of the three-band
model of cuprates has long-range order with a current pattern similar 
to Varma's state.  This claim was brought into question by a subsequent DMRG 
calculation~\cite{nishimoto} on
the same model: the current-current correlations were seen (numerically) 
to decay with a power law.  But that, of course, indicates the presence
of gapless excitations, like the Goldstone mode of a continuous symmetry;
it would {\it not} expected for the Ising-like symmetry of a (pseudo)current
operator, unless the system is accidentally at a critical point. 

In our spinless model, it was easy enough to get current order 
degenerate with bond-density order, or to stabilize the latter,
but quite hard to stabilize just current order.  Could this
be going in the ladder system of Refs.~\onlinecite{new-giamarchi}
and ~\onlinecite{nishimoto}, in which the current operator is just
one component of a larger order parameter with XY symmetry?
We warn, however, that our results also suggest that ladders 
(applied as an approach to 
square lattice models~\cite{schollwock-ladder})
are plain deceptive.  The key terms stabilizing currents depended 
on the very anisotropy of the ladder.

For that reason, it is interesting that Ref.~\onlinecite{weber-giamarchi}
do claim to obtain currents in a variational and truly two-dimensional
calculation.  So far, there is no clear picture of which interactions
are crucial to causing the order.  A strong-coupling caricature 
of the three-band lattice (or ladder) in the spirit of our 
models might illuminate this point.

\SAVE{Sumi agrees that in Ref~\onlinecite{weber-giamarchi}
they don't know what parameters or terms are responsible? }

\SAVE{\subsection{State graph idea}}

\SAVE{Here I refer to the state graph idea mentioned in Zhang and Henley 2003,
and otherwise elaborated only in CLH old proposals (2002 and 2005?).
The state graph reduces a many body system to a single ``particle''
hopping on a very complex graph. 
The key idea is that, to have a real current, we must have one on
the state graph.  If we have a loop on the state graph with a
net $\pi$ factor due to fermion signs, there must either be a
net twist (i.e. a current) or a node.  Any non-uniformity 
is like a transverse field in a two level system and favors
a superposition of the two senses, i.e. a node.
There are a great many systems which (in their state graphs)	
have loops which should give spontaneous currents,
but they do not on account of such transverse-field terms,
similar to the ones we studied.}

\SAVE{Mapped to our spinless language by Nersesyan's map
(see Sec.~\ref{sec:nersesyan}
the model of Tsuchiizu and Furusaki~\cite{furusaki}
contains a correlated double hop,
i.e. a ring exchange, wherein fermions on sites 1 and 3 of the
ring move to 2 and 4 or vice versa.  If we analyze this
from the viewpoint of the ``state graph'', this special
hop produces a triangle of transitions, such that the
net sign of the amplitudes is negative, and thus it
agrees with our scenario for the emergence of spontaneous
currents on the state graph.  However, the net minus sign
cannot be attributed to an odd fermion permutation in this
case: it is impossible to tell which fermion went to which 
destination in the ring exchange, hence its permutation sign 
is undefined.  The net minus sign is instead due to the sign 
of the ring exchange amplitude, which was put in by hand.
We feel the interactions in our model(s) are somewhat more
natural. }

\acknowledgments

C.L.H. thanks S.-A. Cheong, D. I. Khomski,
U. Hizi, M. Troyer, G. Kotliar, and J. B. Marston
for discussions.  This work was supported by NSF grant DMR-0552461.

\appendix

\section{Canonical Transformation}
\label{app:canonical}

Since we use the method of Canonical Transformation for our  calculations, 
here is a very brief summary of the method and the results that are used. 
Say that $\HHO$ represents the strong-coupling limit, which is
assumed to have eigenenergies $\{ E_\alpha \}$ with a large ground
state degeneracy, which however is split by a small perturbation
$\HH'$  (that has no matrix elements between degenerate states of
$\HHO$). We desire to project our problem onto the ground-state
(``zero'') subspace of $\HHO$. The usual way to accomplish this is
canonical transformation: let $\tilde{\HH} \equiv e^{i\Sop}\HH
e^{-i\Sop}$, where we require $\tilde{\HH}_{\alpha\beta} =0$ for
any matrix element connecting the zero subspace to other states;
then, we can restrict our Hilbert space to the span of $e^{i\Sop}
|\alpha\rangle$, where $|\alpha\rangle$ was one of the ground
states.

To lowest order in $\HH'$, the standard canonical transformation is
given by $\Sop_{\alpha \beta}\equiv 0$ when states $\alpha$ and
$\beta$ are degenerate, and $\Sop_{\alpha \beta}\equiv i
\HH'_{\alpha\beta}/(E_\beta-E_\alpha)$ otherwise.  Then the
effective Hamiltonian is given by $\tilde \HH = \HH + \delta \HH$,
where
   \begin{equation}
       \delta \HH_{\alpha\beta} \equiv - \frac {1}{2}
        \sum _\gamma
          \left( \frac{1}{E_\gamma-E_\alpha}
          +  \frac{1}{E_\gamma-E_\beta}\right)
          \HH'_{\alpha \gamma} \HH'_{\gamma\beta}
   \label{eq:Cantran}
   \end{equation}
is the {\it off-diagonal} second-order perturbation correction.
Similarly, an operator $\Oop$ is transformed to $\tilde{\Oop} = \Oop
+ \delta \Oop$, where (to lowest order)
   \begin{equation}
   \delta \Oop_{\alpha\beta} \equiv i
 \sum_\lambda \left(
\frac{\HH'_{\alpha\lambda} \Oop_{\lambda\beta}}{E_\lambda-E_\alpha}
 - \frac{\Oop_{\alpha\lambda} \HH'_{\lambda\gamma} }{E_\lambda-E_\beta}
  \right) .
   \label{eq:optran}
   \end{equation}

For our problem, $\HHop + \HVU$ is the strong coupling limit and $\Hinter$ is the small perturbation.


\begin{thebibliography}{99}
\bibitem{FN-Vterm} The convenience of factoring out the on-plaquette
interaction was the motivation for our unrealistic form of
interaction, Eq.~(\ref{eq:Ham-V}).
\bibitem{FN-symmetries}
Particle-hole symmetry may also be invoked,
but the reflection symmetry is more general:
a $\hat{P}^x$ term is absent even when there is no particle-hole symmetry.
\bibitem{FN-V2}
Making the second-neighbor interaction $V_2$ different from the others
(which is generic even on a square lattice)
would seem to favor CDW order as described by {\it quadratic} term, 
$\propto T_x^2$; but of course for spin-1/2, 
this is just a constant.
\SAVE{
(From an earlier version)
This would not disturb the two-fold degeneracy of the two-fermion states $|2+\ra$
and $|2-\ra$.  Thus, the result would be similar to Eq.~\eqr{eq:TLadder}. 
This is what one would expect since we still have the time reversal and 
reflection symmetry for the ladder.}
\bibitem{FN-spinfull-analysis}
To analyze these cases more formally, we could adapt
the self-consistent approach of Ref.~\onlinecite{Yao_Kivelson}:
take the expectations of the $\vec{T}_i$ operators to create
an effective spin-chain Hamiltonian for the $\vec{S}_i$ operators,
and vice versa.
\bibitem{FN-Yao-outside}
Below $U_c$,
uniform occupation of the plaquette [see end of Sec.~\ref{sec:oneplaq}]
breaks down; above $U_t$ the angular momentum $\tilde 0$ state is
no longer double-filled due to a Hund's rule-like coupling.
\end{thebibliography}
\end{document}